\documentclass[aip,amsmath, reprint]{revtex4-1}
\usepackage{graphicx}

\begin{document}
\title{Multifunctional Devices and Logic Gates With Undoped Silicon Nanowires}
\author{Massimo Mongillo}
\altaffiliation{Present address: IMEP-LAHC, Grenoble-INP, MINATEC, 3 Parvis Louis N\'eel, B.P. 257, 38016 Grenoble, France, E-mail: massimo.mongillo@gmail.com}
\affiliation{SPSMS/LaTEQS, CEA-INAC/UJF-Grenoble 1, 17 Rue des Martyrs, 38054 Grenoble Cedex 9, France}
\author{Panayotis Spathis}
\altaffiliation{Present address: Institut N\'eel, CNRS and Universit\'e Joseph Fourier, B.P. 166, 38042 Grenoble, France}
\affiliation{SPSMS/LaTEQS, CEA-INAC/UJF-Grenoble 1, 17 Rue des Martyrs, 38054 Grenoble Cedex 9, France}
\author{Georgios Katsaros}
\affiliation{SPSMS/LaTEQS, CEA-INAC/UJF-Grenoble 1, 17 Rue des Martyrs, 38054 Grenoble Cedex 9, France}
\author{Pascal Gentile}
\affiliation{SP2M/SINAPS, CEA-INAC/UJF-Grenoble 1, 17 Rue des Martyrs, 38054 Grenoble Cedex 9, France}
\author{Silvano De Franceschi}
\affiliation{SPSMS/LaTEQS, CEA-INAC/UJF-Grenoble 1, 17 Rue des Martyrs, 38054 Grenoble Cedex 9, France}
\email{silvano.defranceschi@cea.fr}

\begin{abstract}
We report on the electronic transport properties of multiple-gate devices fabricated from undoped silicon nanowires. 
Understanding and control of the relevant transport mechanisms was achieved by means of local electrostatic gating and temperature dependent measurements. The roles of the source/drain contacts and of the silicon channel could be independently evaluated and tuned. 
Wrap gates surrounding the silicide-silicon contact interfaces were proved to be effective in inducing a full suppression of the contact Schottky barriers, thereby enabling carrier injection down to liquid-helium temperature. By independently tuning the effective Schottky barrier heights, a variety of reconfigurable device functionalities could be obtained. 
In particular, the same nanowire device could be configured to work as a Schottky barrier transistor, a Schottky diode or a p-n diode with tunable polarities. This versatility was eventually exploited to realize a NAND logic gate with gain well above one.
\end{abstract}

\maketitle


Nanometer-scale electronic devices fabricated from silicon nanowires (SiNWs) are drawing  significant attention in view of their potential application in  electronics \cite{McAlpine2003}, optoelectronics \cite{Yang2006} and biochemical sensing \cite{Cui2001,Patolsky2006}. The transport properties and the functionality of such electronic devices  are usually controlled by doping. In most cases, the incorporation of doping impurities in SiNWs is obtained \textsl{in-situ} during nanowire growth \cite{Yang2005}, but a precise control over their spatial distribution \cite{Perea2009,Koren2010} and their activation \cite{Bjork2009} has not been achieved yet. Doping control becomes a particularly critical issue when the characteristic device size  approaches the nanometer scale, \textit{i.e.} comparable to the typical distance between dopants for standard doping levels ($10^{17}$ - $10^{19}$ cm$^{-3}$). In this limit, device performances can depend on only a few dopants \cite{Asenov2003,PierreM.2010}, and  be extremely sensitive to their precise locations, leading to a significant device-to-device variability. The main obstacle coming from the use of undoped nanowires lies in the difficulty to form low-resistance contacts due to the unavoidable presence of a Schottky barrier (SB) \cite{Tung2001} at the metal-silicon interface.  Therefore, understanding and controlling the properties of electrical contacts to SiNWs is of fundamental importance \cite{Leonard2011}. Here, we investigate the properties of metal-silicide contacts to undoped SiNWs, and we study the possibility to obtain largely-tunable contact resistances through a combination of two fabrication processes: a controlled silicidation of the metal-SiNW contacts and the fabrication of local gate electrodes wrapped around each silicon-silicide interface. We demonstrate that contact resistances can be largely suppressed with gate voltages of the order of 1 V, enabling measurable carrier injection down to 4K. In addition, through local electrostatic doping of the SiNW, our  approach provides the possibility to implement different functionalities within the same SiNW device, which can work as a bipolar transistor, a Schottky diode or p-n diode with gate-tunable polarities. Finally, we provide as well an example of how two such devices could be programmed to operate as a NAND logic gate.

We used undoped SiNWs grown by chemical vapor deposition via a catalytic vapor-liquid-solid method  (growth details were given in an earlier work \cite{Gentile2008}). Our sample fabrication process relies on a few steps of e-beam lithography, e-beam metal deposition, and lift-off. Initially, alignment markers and ordered arrays of 500-nm-wide Cr/Al (10/5 nm) gate electrodes, spaced by 500 nm, are defined on top of an oxidized heavily doped silicon substrate ($p^{++}$ $Si/SiO_{2}$ ). Successively, after a brief sonication in isopropanol, the as-grown undoped SiNWs are released from their growth substrate and drop-casted on the device substrate. Wires crossing a couple of bottom gates are identified by means of scanning electron microscopy (SEM) and contacted with 120-nm-thick nickel electrodes to be used as source and drain contacts. The bottom gates are also contacted in the same lithographic step. 
Prior to the Ni deposition, the native oxide on the SiNWs is etched away by dipping the sample for 5s in buffered HF. This step removes also the Al cap layer of the bottom gates leaving the underlying Cr layer unaffected. After contacting the SiNWs  a double step thermal annealing process is performed in order to promote the silicidation of the contacts. Temperature and annealing times are calibrated in order obtain the protrusion of the nickel silicide along the nanowire \cite{Weber2006} from the Ni contacts to the bottom gates. The preparation of the sample is completed with the deposition of 80-nm-thick Al top gates overlapping with the bottom gates and, simultaneously, an additional top gate electrode crossing the middle region of the SiNW. The Al top gates are electrically isolated from the nanowire by a preliminary deposition and oxidation of four 1.5-nm-thick Al layers. Each layer is fully oxidized by leaving the sample in the load-lock chamber of the e-beam evaporator under an oxygen pressure of 200 Torr for 10 minutes.  
Due to their  large overlapping area and to  the small thickness of the (native) oxide interlayer,  each bottom gate	and the corresponding  top gate are effectively shorted together, resulting in a wrap-gate geometry \cite{Storm2011}. 
\begin{figure}[h]
\begin{center}
\includegraphics[width=8.65cm,keepaspectratio]{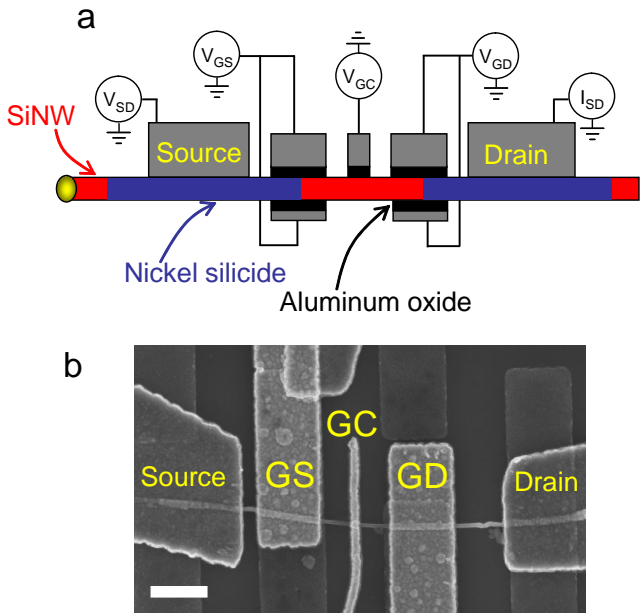}
\caption{a) Schematic of a multi-gate device made from a single, undoped SiNW. Two wrap-gates, labeled as GS and GD, are designed to control the Schottky barriers at the silicide-silicon junctions formed by the source and drain contacts. The finger gate in the middle, labeled as GC, is meant to control carrier population in the silicon channel. b) SEM micrograph of the device. Scale bar: 400nm.
}
\label{fig1}
\end{center}
\end{figure}

The SiNW device and its schematics are shown in \ref{fig1}. Since the nickel silicide sections have a metallic character, SB contacts are formed at the silicon-silicide interfaces. The wrap-gate electrodes, labelled as GS and GD,  are designed to embed these interfaces and to control the respective SBs.  We shall refer to these electrodes as to the  \textit{contact gates}. The wrap-gate geometry maximizes the gating efficiency. The finger-shape gate, labelled as GC, is meant to control carrier population in the  channel. We shall refer to this electrode as to the \textit{channel gate}.

\begin{figure}[h]
\begin{center}
\includegraphics[width=8.65cm,keepaspectratio]{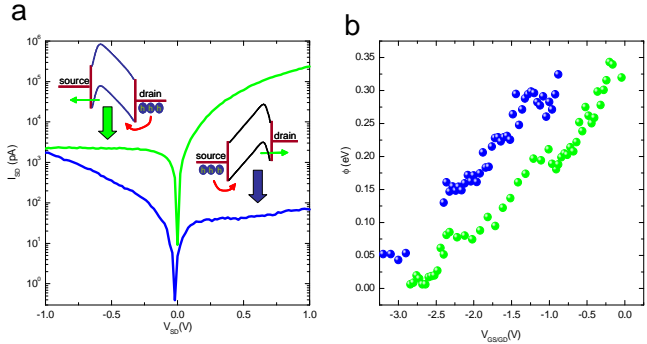}
\caption{
Schottky-diode $I_{SD}(V_{SD})$ characteristics obtained for two gate settings. The blue trace is obtained with $V_{GD}=-3.2$ V and $V_{GS}=0$. In this configuration, the drain effective  SB is suppressed while the source SB is left unaltered, thereby leading to rectifying behaviour. The green trace is obtained with $V_{GS}=-3.2$ V and $V_{GD}=0$, resulting in an opposite rectifying behavior. Insets: schematic band diagrams illustrating the thermal emission of carriers over the reverse-bias Schottky contact.
b) Contact-gate dependence of the effective (p-type) SB height for the source (blue symbols) and the drain (green symbols) contacts. 
The horizontal axis refers to $V_{GS}$ and $V_{GD}$, respectively.  
}
\label{fig2}
\end{center}
\end{figure}

The multiple gate layout enables independent control of the different device sections, \textit{i.e.} the SB contacts and the channel.  
We shall begin by evaluating their relative contributions under different gating conditions and, for each case, we shall investigate the relevant transport mechanisms. At room temperature and for zero gate voltages, thermionic emission over the contact SBs is the dominant transport process. Since the Fermi level of the nickel silicide is pinned below the silicon mid-gap, \textit{i.e.} closer to the valence band, transport across each silicide-silicon junction is dominated by the thermionic emission of holes over the p-type SB. The application of negative voltages to the  contact-gates GS/GD produces a local band bending that reduces the thickness of the SB of the corresponding
silicide-silicon junction. This enables the onset of thermally assisted tunneling through the barrier top-edge leading to  a reduction of the effective p-type SB height 
 \cite{Appenzeller2004a,Yang2008}, $\phi$. We find that negative contact-gate voltages of a few V are sufficient to induce a strong suppression of $\phi $. This is apparent from \ref{fig2}(a) where the source-drain current, $I_{SD} $, is plotted as a function of the source-drain bias voltage, $V_{SD} $, for two gating conditions. The green characteristic corresponds to $V_{GS}=-3.2$ V and $V_{GD}=0$. In this case, the effective SB at the source contact is suppressed and the device behaves as a Schottky diode whose polarity is imposed  by the SB at the drain contact. The blue characteristic corresponds to the reversed gating condition, i.e. $V_{GD}=-3.2$ V and $V_{GS}=0$, which results in an opposite diode polarity. In both  cases the channel gate was set to $-2.5$ V in order to get rid of any possible barrier in the channel (see discussion further below). 
Appreciable quantitative discrepancies can be noted in the two characteristics of \ref{fig2}(a).
Under reverse-bias polarization, described by the qualitative band diagrams in the two insets of \ref{fig2}(a), the measured current is dominated by the thermally assisted tunneling of holes across the reverse-bias SB, \textit{i.e.}   
\begin{equation}
I_{SD}= A^{\star}ST^{2}exp(-\frac{e\phi}{k_{B}T}) 
\label{equation Arrhenius}
\end{equation}
where $k_{B}= 8.617$eV/K is the Boltzmann constant, $A^{\star}$ is the Richardson  constant, $T $ is the temperature, $e $ is the absolute value of the electronic charge, $S $ is the cross-sectional area of the conducting silicon channel.
From a ratio of $\sim 50$ between the reverse currents of the green and the blue characteristics, we estimate that the drain SB height exceeds the source SB height by $k_BT ln(50) \approx 0.1 eV$.
   

In order to independently measure the two effective SB heights and to study their gate-voltage dependence, the sample was loaded in a home-made variable-temperature insert.  For each gate-voltage setting, temperature was varied between 220 K and 370 K in steps of 10 K. The effective barrier height was extracted from the temperature dependence of the reverse-bias current using the thermionic-emission relation (\ref{equation Arrhenius}) ($\phi$ is given by the the slope of the Arrhenius plot $\ln({I_{SD}/T^{2}})$ vs $1/T$).
The results are shown in \ref{fig2}(b), where the effective SB heights are plotted as a function of the corresponding contact-gate voltages.
The two data sets reveal approximately linear gate-voltage dependences with same  slopes. This denotes almost identical capacitive couplings to the respective contact gates. 
The drain SB height is 0.35 eV for $V_{GD} = 0$ and it vanishes for $V_{GD} = -2.8$V. The source SB height for $V_{GD} = 0$ could not be directly measured due to the limited upper temperature of our setup. Yet a linear extrapolation of the data in \ref{fig2}(b) yields a source SB height of 0.45 eV for $V_{GD} \rightarrow 0$. This confirms the 0.1 eV difference with the drain SB height previously deduced from the ratio of the reverse currents in \ref{fig2}(a). Because of this difference, the source SB requires slightly more negative gate voltages to be entirely suppressed. 

\begin{figure}[h]
\begin{center}
\includegraphics[width=8.65cm,keepaspectratio]{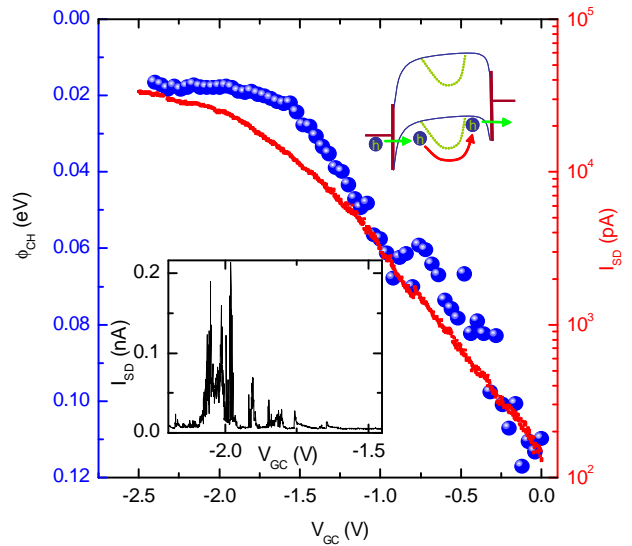}
\caption{The effective  SBs of the source and drain contacts are simultaneously suppressed by setting $V_{GS}=V_{GD}=-3.2$ V. As a result, the device can be operated as a p-type field-effect transistor, using the channel-gate voltage, $V_{GC}$, to modulate the source-drain current, $I_{SD}$.  Red trace: transfer characteristic, $I_{SD}(V_{GC})$, taken with $V_{SD}=100$ mV at room temperature. Blue symbols: channel barrier height, $\phi_{CH}$, as a function of $V_{GC}$. Upper inset: schematic band diagram illustrating the thermal emission of holes over the gate-controlled barrier in the middle region of the silicon channel (green dotted line). Lower inset: $I_{SD}(V_{GC})$ characteristic taken for $V_{SD}=100$ mV at $T=4$K.
 }
\label{fig3}
\end{center}
\end{figure}

If both the effective SBs are simultaneously suppressed by sufficiently negative voltages applied to the respective contact gates, transport across the device becomes entirely dependent on the hole population of the silicon channel \cite{Appenzeller2004}. We find that current transport is dominated by the thermionic emission of holes over a potential barrier formed by the downward bending of the valence band edge. The emission of carriers over this barrier is purely thermionic   as opposed to the  case of the contact SBs where the onset of tunnelling leads to a smaller effective barrier. Following a similar procedure as before, we have measured the height of this barrier as a function of  
$V_{GC}$. The results are shown by solid dots in \ref{fig3}, where each dot is the result of a fit to an Arrhenius plot.

All data points were taken at constant $V_{SD} = -1$ V. A room-temperature $I_{SD}(V_{GC})$ characteristic, taken in the same $V_{GC}$ range, is also shown on a logarithmic scale (red trace). The two data sets meet fairly well the proportionality relation expected from the thermionic-emission relation (\ref{equation Arrhenius}). The measured barrier height is $\sim0.11$ eV for  $V_{GC} = 0$ V and it decreases linearly for negative gate voltages. Below $V_{GC} < - 1.5$ V, the barrier becomes too small to be reliably extracted from the Arrhenius plot (in this regime, transport is no longer thermally activated). 
A saturation of the $I_{SD}(V_{GC})$ characteristic is consistently observed below this threshold voltage. 

The complete suppression of the silicon-channel barrier, together with the suppression of the effective SBs at the silicon-silicide contacts, are further confirmed by low temperature measurements. 
An $I_{SD}(V_{GC})$ characteristic taken at $T=4$ K for $V_{SD}=0.1$ V is shown in the lower inset of \ref{fig3}. While $I_{SD}$ is unmeasurably small for $V_{GC} > -1.6$ V, pronounced current oscillations can be seen below this threshold, providing evidence of low-temperature transport through an undoped silicon device. The observed current oscillations are due to the Coulomb blockade effect. Their irregular structure reflects single-hole tunneling through a series of hole islands formed along the nanowire as it is typically observed in disordered low-dimensional conductors.


So far we have shown full control of hole transport in an undoped SiNW device. The same device could be operated as a p-type field-effect transistor, a Schottky diode, or a multi-island single-hole tunneling device. We pointed out that the p-type character of all these functionalities arises from the pinning of the silicide Fermi energy closer to the silicon valence band edge. At the same time, our thermionic-emission measurements revealed that the SB height can exhibit appreciable contact-to-contact variations within the same device.
These variations can arise from the atomic structure of the silicide-silicon junction \cite{Gambino1998,Iwai2002,Morimoto1995,Murarka1995,Bucher1986-06-01} as well as from the field effect of trapped charges in the periphery of the silicide-silicon junction \cite{Leonard2011,Cai2012a}.   

\begin{figure}[h]
\begin{center}
\includegraphics[width=8.65cm,keepaspectratio]{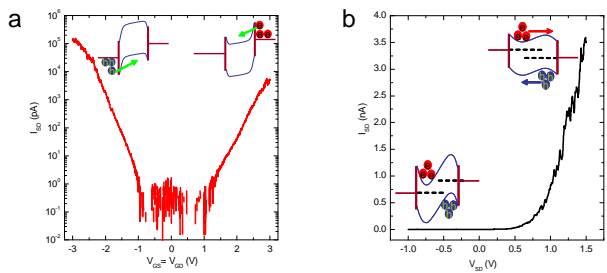}
\caption{ a) Room-temperature transfer characteristic of a dual-gate device showing bipolar transistor behavior. Hole-dominated (electron-dominated) transport is obseved for negative (positive) gate voltages, as schematically shown in the left (right) inset. This measurement was taken with $V_{SD}=0.5$ V and $V_{GS}=V_{GD}$. b) $I_{SD}(V_{SD})$ characteristic showing the rectifying behavior expected for an electrically doped p-n diode. The tunable "doping" is achieved with $V_{GS}= 2$ V and $V_{GD}= -2$ V. Schematic band diagrams for forward- and reverse-bias conditions are shown in the upper and lower insets, respectively.}
\label{fig4}
\end{center}
\end{figure}

Given the relatively large variability ($\sim 0.1$ eV) in the SB height of silicon-silicide contacts, it is possible to have SiNW devices where Fermi level pinning occurs close to the silicon mid-gap, yielding p-type and n-type SBs relatively close to each other and to the half-gap value (0.55 eV). These types of devices exhibit bipolar behavior and, as a result, they can give rise to a larger range of functionalities. 
\ref{fig4}(a) shows the bipolar field-effect characteristic of one of such devices with a gate layout identical to that of \ref{fig1}, except for the absence in this case of the channel-gate electrode. The horizontal axis refers to both contact-gate voltages being swept together. 

A more interesting behaviour is observed when the two gates are polarized at opposite voltages, fixed with respect to the corresponding (source or drain) contacts. 
\ref{fig4}(b) shows that, in this case, the device behaves like an electrostatically doped \textsl{p-n} diode. 
As opposed to the case of true $p-n$ diodes, here the built-in potential arises from static charges accumulated on the contact-gate electrodes \cite{Lee2004,Mueller2010,Heinzig2011,Martin2011} and not on the potential of the ionized impurities in the space-charge region. 
In the forward bias polarization the side of the nanowire populated with holes has a higher electrochemical potential than  the side populated by electrons. The barriers created  by the contact gate become smaller and can be surmounted by thermally activated  carriers as shown in the upper inset of \ref{fig4}(b). 
This results in a diffusion current of holes from the \textsl{p}-like to the \textsl{n}-like region and an electron current in the opposite direction. In the case of  reverse bias polarization, transport of electrons and holes is hindered by the high potential barriers created by the contact gates as depicted in the lower inset of \ref{fig4}(b). 

\begin{figure}[h]
\begin{center}
\includegraphics[width=8.5cm,keepaspectratio]{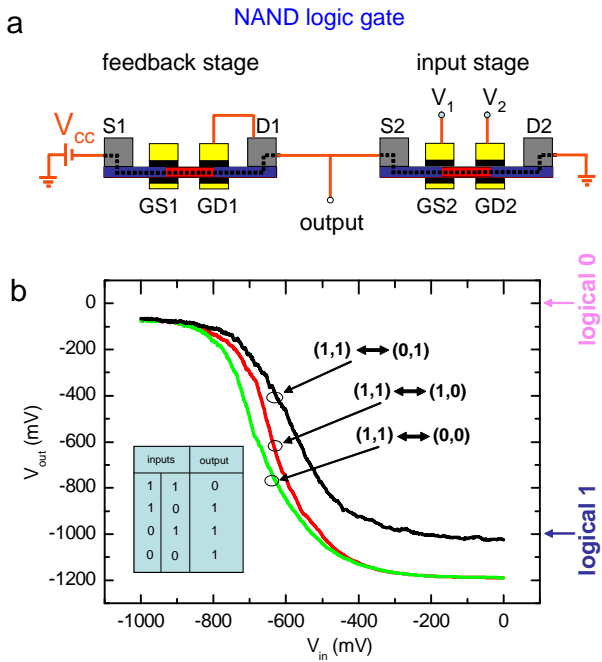}
\caption{a) Schematic of a NAND logic gate based on two dual-gate SiNW devices. One device acts as an input stage, while the second one acts a feedback stage to enforce the correct output voltage. The two input lines are fed to the contact gates GS2 and GD2. The output line is taken from the D1 contact, which is shorted with the S2 contact. The feedback function is accomplished by applying a fixed negative polarization of $-3$ V to GS1, large enough to suppress the corresponding effective SB, and by connecting GD1 to the output terminal.  A voltage $V_{cc} = - 1.2$ V is applied to S1, with D2 at ground.  b) Output characteristics of the logic gate. The black trace, obtained with  $V_{in}=V_{1}$ and $V_{2}=-1$ V, corresponds to the transition  $(1,1) \leftrightarrow (0,1)$ in the input levels. The red trace, obtained with $V_{in}=V_{2} $ and $V_{1}=-1$ V, corresponds to the transistion  $(1,1) \leftrightarrow (1,0)$. The green trace, obtained with $V_{in}=V_{1}=V_{2}$, corresponds to the transistion $(1,1) \leftrightarrow (0,0)$. Inset: truth table for a NAND logic gate.}
\label{fig5}
\end{center}
\end{figure}

Below we show that dual-gate SiNW devices, having Schottky contacts well coupled to the respective gates, can serve as building blocks for logic circuits. Bottom-up logic gates based on carbon nanotubes \cite{Bachtold2001,Derycke2001} or semiconductor nanowires \cite{Cui2001,Huang2001} have already been reported,  along with more complex circuits like ring oscillators \cite{Chen2006} extending digital operation into the high frequency domain. 
Here we demonstrate the implementation of a NAND device made from two dual-gate SiNW devices connected in series as in \ref{fig5}(a). 
Since the SiNWs are undoped, this approach has the advantage of not requiring complementary doping.
A NAND device performs an AND and a NOT operation in series. It accepts two input voltages corresponding to the binary values $0$ and $1$ and it delivers a single output voltage according to the truth table in \ref{fig5}(b). In our circuit, the voltage level $V=0$V corresponds to a logical $0$, while the voltage level $V=-1$V corresponds to a logical $1$.

Let us now analyze the operating principle of this NAND device, by considering the different input configurations. 
When the two input gates controlling $GS2$ and $GD2$ are at logical $0$, the input stage is in a highly resistive state due to both SBs being simultaneously high. As a consequence the bias voltage, $V_{cc} = -1.2$ V, falls mostly across the feedback stage and the output terminal acquires a value $V_{out} \approx V_{cc}$ corresponding to a logical 1. The presence of an electrical connection between the output terminal and the contact gate $GD1$ produces a positive feedback, since a negative value on the contact gate suppresses the corresponding effective  SB, making the impedance of the feedback stage lower and hence forcing $V_{out} $ even closer to $V_{cc}$, i.e. to logical output 1. When, on the contrary, both inputs are at logical $1$, the input stage  attains a low-resistive state due to the simultaneous suppression of both effective SBs. As a result, the output voltage approaches a logical 0, and $V_{cc}$ falls mainly  on the feeback stage. Because of the feedback line, $V_{out}\approx 0 $V enforces a high value of the SB at the drain contact of the feedback stage. This increases its resistance with respect to the input stage, thereby reinforcing logical output 0.  In the intermediate regimes, when one of the two inputs is at logical $0 $ and the other is at logical $1 $, one gated SB is suppressed while the other is not. In this case, the presence of the feedback line is crucial in producing a stable configuration with $V_{out}\approx V_{cc}$, i.e. logical output 1. In fact, under this condition both of the effective  SBs at the source and drain contacts of the feedback stage are suppressed leading to a lower impedance  as compared to the input stage. As a result, the output voltage is stabilized at a logical 1. 

Since in actual logic circuits the output of a logic gate can serve as the input of another gate, an important figure of merit for logic gates is gain, $g$, defined as the slope of the linear portion of the output characteristic. A high-gain logic gate ($g > 1$) has the capability to drive another logic gate without the need of signal restoration. From the data in \ref{fig5}(b) we find $g$ between $2.6 $ and $4.7$. In addition, the NAND gate requires relatively low voltage levels as compared to previously reported logic gates built from semiconductor nanowires \cite{Cui2001,Huang2001}. Similar to other earlier works\cite{Bachtold2001}, our NAND device was assembled from two dual-gate devices connected via external leads. Yet we should like to point out that the same logic gate could be obtained with a single nanowire using a common electrode for D1 and S2.
   
We thank the technical staff of the PTA cleanroom,  G.Lapertot and C. Marin for their help in device fabrication
and technical support. Massimo Mongillo gratefully acknowledges Antonella Orsino for useful discussions. This work was supported by the Agence Nationale de la Recherche (ANR) through the ACCESS and COHESION projects and by
the European Commission through the Chemtronics program MEST-CT-2005-020513.

\bibliographystyle{apsrev}

\bibliography{bibliography_nanowires}

\begin{thebibliography}{32}
\expandafter\ifx\csname natexlab\endcsname\relax\def\natexlab#1{#1}\fi
\expandafter\ifx\csname bibnamefont\endcsname\relax
  \def\bibnamefont#1{#1}\fi
\expandafter\ifx\csname bibfnamefont\endcsname\relax
  \def\bibfnamefont#1{#1}\fi
\expandafter\ifx\csname citenamefont\endcsname\relax
  \def\citenamefont#1{#1}\fi
\expandafter\ifx\csname url\endcsname\relax
  \def\url#1{\texttt{#1}}\fi
\expandafter\ifx\csname urlprefix\endcsname\relax\def\urlprefix{URL }\fi
\providecommand{\bibinfo}[2]{#2}
\providecommand{\eprint}[2][]{\url{#2}}

\bibitem[{\citenamefont{McAlpine et~al.}(2003)\citenamefont{McAlpine, Friedman,
  Jin, Lin, Wang, and Lieber}}]{McAlpine2003}
\bibinfo{author}{\bibfnamefont{M.~C.} \bibnamefont{McAlpine}},
  \bibinfo{author}{\bibfnamefont{R.~S.} \bibnamefont{Friedman}},
  \bibinfo{author}{\bibfnamefont{S.}~\bibnamefont{Jin}},
  \bibinfo{author}{\bibfnamefont{K.-h.} \bibnamefont{Lin}},
  \bibinfo{author}{\bibfnamefont{W.~U.} \bibnamefont{Wang}}, \bibnamefont{and}
  \bibinfo{author}{\bibfnamefont{C.~M.} \bibnamefont{Lieber}},
  \bibinfo{journal}{Nano Letters} \textbf{\bibinfo{volume}{3}},
  \bibinfo{pages}{1531} (\bibinfo{year}{2003}), ISSN \bibinfo{issn}{1530-6984}.

\bibitem[{\citenamefont{Yang et~al.}(2006)\citenamefont{Yang, Barrelet,
  Capasso, and Lieber}}]{Yang2006}
\bibinfo{author}{\bibfnamefont{C.}~\bibnamefont{Yang}},
  \bibinfo{author}{\bibfnamefont{C.~J.} \bibnamefont{Barrelet}},
  \bibinfo{author}{\bibfnamefont{F.}~\bibnamefont{Capasso}}, \bibnamefont{and}
  \bibinfo{author}{\bibfnamefont{C.~M.} \bibnamefont{Lieber}},
  \bibinfo{journal}{Nano Lett.} \textbf{\bibinfo{volume}{6}},
  \bibinfo{pages}{2929} (\bibinfo{year}{2006}), ISSN \bibinfo{issn}{1530-6984}.

\bibitem[{\citenamefont{Cui and Lieber}(2001)}]{Cui2001}
\bibinfo{author}{\bibfnamefont{Y.}~\bibnamefont{Cui}} \bibnamefont{and}
  \bibinfo{author}{\bibfnamefont{C.~M.} \bibnamefont{Lieber}},
  \bibinfo{journal}{Science} \textbf{\bibinfo{volume}{291}},
  \bibinfo{pages}{851} (\bibinfo{year}{2001}).

\bibitem[{\citenamefont{Patolsky et~al.}(2006)\citenamefont{Patolsky, Timko,
  Yu, Fang, Greytak, Zheng, and Lieber}}]{Patolsky2006}
\bibinfo{author}{\bibfnamefont{F.}~\bibnamefont{Patolsky}},
  \bibinfo{author}{\bibfnamefont{B.~P.} \bibnamefont{Timko}},
  \bibinfo{author}{\bibfnamefont{G.}~\bibnamefont{Yu}},
  \bibinfo{author}{\bibfnamefont{Y.}~\bibnamefont{Fang}},
  \bibinfo{author}{\bibfnamefont{A.~B.} \bibnamefont{Greytak}},
  \bibinfo{author}{\bibfnamefont{G.}~\bibnamefont{Zheng}}, \bibnamefont{and}
  \bibinfo{author}{\bibfnamefont{C.~M.} \bibnamefont{Lieber}},
  \bibinfo{journal}{Science} \textbf{\bibinfo{volume}{313}},
  \bibinfo{pages}{1100} (\bibinfo{year}{2006}).

\bibitem[{\citenamefont{Yang et~al.}(2005)\citenamefont{Yang, Zhong, and
  Lieber}}]{Yang2005}
\bibinfo{author}{\bibfnamefont{C.}~\bibnamefont{Yang}},
  \bibinfo{author}{\bibfnamefont{Z.}~\bibnamefont{Zhong}}, \bibnamefont{and}
  \bibinfo{author}{\bibfnamefont{C.~M.} \bibnamefont{Lieber}},
  \bibinfo{journal}{Science} \textbf{\bibinfo{volume}{310}},
  \bibinfo{pages}{1304} (\bibinfo{year}{2005}).

\bibitem[{\citenamefont{Perea et~al.}(2009)\citenamefont{Perea, Hemesath,
  Schwalbach, Lensch-Falk, Voorhees, and Lauhon}}]{Perea2009}
\bibinfo{author}{\bibfnamefont{D.~E.} \bibnamefont{Perea}},
  \bibinfo{author}{\bibfnamefont{E.~R.} \bibnamefont{Hemesath}},
  \bibinfo{author}{\bibfnamefont{E.~J.} \bibnamefont{Schwalbach}},
  \bibinfo{author}{\bibfnamefont{J.~L.} \bibnamefont{Lensch-Falk}},
  \bibinfo{author}{\bibfnamefont{P.~W.} \bibnamefont{Voorhees}},
  \bibnamefont{and} \bibinfo{author}{\bibfnamefont{L.~J.}
  \bibnamefont{Lauhon}}, \bibinfo{journal}{Nat Nano}
  \textbf{\bibinfo{volume}{4}}, \bibinfo{pages}{315} (\bibinfo{year}{2009}),
  ISSN \bibinfo{issn}{1748-3387}.

\bibitem[{\citenamefont{Koren et~al.}(2010)\citenamefont{Koren, Berkovitch, and
  Rosenwaks}}]{Koren2010}
\bibinfo{author}{\bibfnamefont{E.}~\bibnamefont{Koren}},
  \bibinfo{author}{\bibfnamefont{N.}~\bibnamefont{Berkovitch}},
  \bibnamefont{and}
  \bibinfo{author}{\bibfnamefont{Y.}~\bibnamefont{Rosenwaks}},
  \bibinfo{journal}{Nano Lett.} \textbf{\bibinfo{volume}{10}},
  \bibinfo{pages}{1163} (\bibinfo{year}{2010}), ISSN \bibinfo{issn}{1530-6984}.

\bibitem[{\citenamefont{Bjork et~al.}(2009)\citenamefont{Bjork, Schmid, Knoch,
  Riel, and Riess}}]{Bjork2009}
\bibinfo{author}{\bibfnamefont{M.~T.} \bibnamefont{Bjork}},
  \bibinfo{author}{\bibfnamefont{H.}~\bibnamefont{Schmid}},
  \bibinfo{author}{\bibfnamefont{J.}~\bibnamefont{Knoch}},
  \bibinfo{author}{\bibfnamefont{H.}~\bibnamefont{Riel}}, \bibnamefont{and}
  \bibinfo{author}{\bibfnamefont{W.}~\bibnamefont{Riess}},
  \bibinfo{journal}{Nat Nano} \textbf{\bibinfo{volume}{4}},
  \bibinfo{pages}{103} (\bibinfo{year}{2009}), ISSN \bibinfo{issn}{1748-3387}.

\bibitem[{\citenamefont{Asenov et~al.}(2003)\citenamefont{Asenov, Brown,
  Davies, Kaya, and Slavcheva}}]{Asenov2003}
\bibinfo{author}{\bibfnamefont{A.}~\bibnamefont{Asenov}},
  \bibinfo{author}{\bibfnamefont{A.}~\bibnamefont{Brown}},
  \bibinfo{author}{\bibfnamefont{J.}~\bibnamefont{Davies}},
  \bibinfo{author}{\bibfnamefont{S.}~\bibnamefont{Kaya}}, \bibnamefont{and}
  \bibinfo{author}{\bibfnamefont{G.}~\bibnamefont{Slavcheva}},
  \bibinfo{journal}{Electron Devices, IEEE Transactions}
  \textbf{\bibinfo{volume}{50}}, \bibinfo{pages}{1837} (\bibinfo{year}{2003}),
  ISSN \bibinfo{issn}{0018-9383}.

\bibitem[{\citenamefont{PierreM. et~al.}(2010)\citenamefont{PierreM.,
  WacquezR., JehlX., SanquerM., VinetM., and CuetoO.}}]{PierreM.2010}
\bibinfo{author}{\bibnamefont{PierreM.}},
  \bibinfo{author}{\bibnamefont{WacquezR.}},
  \bibinfo{author}{\bibnamefont{JehlX.}},
  \bibinfo{author}{\bibnamefont{SanquerM.}},
  \bibinfo{author}{\bibnamefont{VinetM.}}, \bibnamefont{and}
  \bibinfo{author}{\bibnamefont{CuetoO.}}, \bibinfo{journal}{Nat Nano}
  \textbf{\bibinfo{volume}{5}}, \bibinfo{pages}{133} (\bibinfo{year}{2010}).

\bibitem[{\citenamefont{Tung}(2001)}]{Tung2001}
\bibinfo{author}{\bibfnamefont{R.~T.} \bibnamefont{Tung}},
  \bibinfo{journal}{Materials Science and Engineering: R: Reports}
  \textbf{\bibinfo{volume}{35}}, \bibinfo{pages}{1} (\bibinfo{year}{2001}),
  ISSN \bibinfo{issn}{0927-796X}.

\bibitem[{\citenamefont{Leonard and Talin}(2011)}]{Leonard2011}
\bibinfo{author}{\bibfnamefont{F.}~\bibnamefont{Leonard}} \bibnamefont{and}
  \bibinfo{author}{\bibfnamefont{A.~A.} \bibnamefont{Talin}},
  \bibinfo{journal}{Nat Nano} \textbf{\bibinfo{volume}{6}},
  \bibinfo{pages}{773} (\bibinfo{year}{2011}), ISSN \bibinfo{issn}{1748-3387}.

\bibitem[{\citenamefont{Gentile et~al.}(2008)\citenamefont{Gentile, David,
  Dhalluin, Buttard, Pauc, Den~Hertog, Ferret, and Baron}}]{Gentile2008}
\bibinfo{author}{\bibfnamefont{P.}~\bibnamefont{Gentile}},
  \bibinfo{author}{\bibfnamefont{T.}~\bibnamefont{David}},
  \bibinfo{author}{\bibfnamefont{F.}~\bibnamefont{Dhalluin}},
  \bibinfo{author}{\bibfnamefont{D.}~\bibnamefont{Buttard}},
  \bibinfo{author}{\bibfnamefont{N.}~\bibnamefont{Pauc}},
  \bibinfo{author}{\bibfnamefont{M.}~\bibnamefont{Den~Hertog}},
  \bibinfo{author}{\bibfnamefont{P.}~\bibnamefont{Ferret}}, \bibnamefont{and}
  \bibinfo{author}{\bibfnamefont{T.}~\bibnamefont{Baron}},
  \bibinfo{journal}{Nanotechnology} \textbf{\bibinfo{volume}{19}},
  \bibinfo{pages}{125608} (\bibinfo{year}{2008}), ISSN
  \bibinfo{issn}{0957-4484}.

\bibitem[{\citenamefont{Weber et~al.}(2006)\citenamefont{Weber, Geelhaar,
  Graham, Unger, Duesberg, Liebau, Pamler, Cheze, Riechert, Lugli
  et~al.}}]{Weber2006}
\bibinfo{author}{\bibfnamefont{W.~M.} \bibnamefont{Weber}},
  \bibinfo{author}{\bibfnamefont{L.}~\bibnamefont{Geelhaar}},
  \bibinfo{author}{\bibfnamefont{A.~P.} \bibnamefont{Graham}},
  \bibinfo{author}{\bibfnamefont{E.}~\bibnamefont{Unger}},
  \bibinfo{author}{\bibfnamefont{G.~S.} \bibnamefont{Duesberg}},
  \bibinfo{author}{\bibfnamefont{M.}~\bibnamefont{Liebau}},
  \bibinfo{author}{\bibfnamefont{W.}~\bibnamefont{Pamler}},
  \bibinfo{author}{\bibfnamefont{C.}~\bibnamefont{Cheze}},
  \bibinfo{author}{\bibfnamefont{H.}~\bibnamefont{Riechert}},
  \bibinfo{author}{\bibfnamefont{P.}~\bibnamefont{Lugli}},
  \bibnamefont{et~al.}, \bibinfo{journal}{Nano Letters}
  \textbf{\bibinfo{volume}{6}}, \bibinfo{pages}{2660} (\bibinfo{year}{2006}),
  ISSN \bibinfo{issn}{1530-6984}.

\bibitem[{\citenamefont{Storm et~al.}(2011)\citenamefont{Storm, Nylund,
  Samuelson, and Micolich}}]{Storm2011}
\bibinfo{author}{\bibfnamefont{K.}~\bibnamefont{Storm}},
  \bibinfo{author}{\bibfnamefont{G.}~\bibnamefont{Nylund}},
  \bibinfo{author}{\bibfnamefont{L.}~\bibnamefont{Samuelson}},
  \bibnamefont{and} \bibinfo{author}{\bibfnamefont{A.~P.}
  \bibnamefont{Micolich}}, \bibinfo{journal}{Nano Lett.}
  \textbf{\bibinfo{volume}{12}}, \bibinfo{pages}{1} (\bibinfo{year}{2011}),
  ISSN \bibinfo{issn}{1530-6984}.

\bibitem[{\citenamefont{Appenzeller
  et~al.}(2004{\natexlab{a}})\citenamefont{Appenzeller, Radosavljevic, Knoch,
  and Avouris}}]{Appenzeller2004a}
\bibinfo{author}{\bibfnamefont{J.}~\bibnamefont{Appenzeller}},
  \bibinfo{author}{\bibfnamefont{M.}~\bibnamefont{Radosavljevic}},
  \bibinfo{author}{\bibfnamefont{J.}~\bibnamefont{Knoch}}, \bibnamefont{and}
  \bibinfo{author}{\bibfnamefont{P.}~\bibnamefont{Avouris}},
  \bibinfo{journal}{Phys. Rev. Lett.} \textbf{\bibinfo{volume}{92}},
  \bibinfo{pages}{048301} (\bibinfo{year}{2004}{\natexlab{a}}).

\bibitem[{\citenamefont{Yang et~al.}(2008)\citenamefont{Yang, Lee, Liang,
  Eswar, Sun, and Kwong}}]{Yang2008}
\bibinfo{author}{\bibfnamefont{W.}~\bibnamefont{Yang}},
  \bibinfo{author}{\bibfnamefont{S.}~\bibnamefont{Lee}},
  \bibinfo{author}{\bibfnamefont{G.}~\bibnamefont{Liang}},
  \bibinfo{author}{\bibfnamefont{R.}~\bibnamefont{Eswar}},
  \bibinfo{author}{\bibfnamefont{Z.}~\bibnamefont{Sun}}, \bibnamefont{and}
  \bibinfo{author}{\bibfnamefont{D.}~\bibnamefont{Kwong}},
  \bibinfo{journal}{Nanotechnology, IEEE Transactions on}
  \textbf{\bibinfo{volume}{7}}, \bibinfo{pages}{728} (\bibinfo{year}{2008}),
  ISSN \bibinfo{issn}{1536-125X}.

\bibitem[{\citenamefont{Appenzeller
  et~al.}(2004{\natexlab{b}})\citenamefont{Appenzeller, Lin, Knoch, and
  Avouris}}]{Appenzeller2004}
\bibinfo{author}{\bibfnamefont{J.}~\bibnamefont{Appenzeller}},
  \bibinfo{author}{\bibfnamefont{Y.-M.} \bibnamefont{Lin}},
  \bibinfo{author}{\bibfnamefont{J.}~\bibnamefont{Knoch}}, \bibnamefont{and}
  \bibinfo{author}{\bibfnamefont{P.}~\bibnamefont{Avouris}},
  \bibinfo{journal}{Phys. Rev. Lett.} \textbf{\bibinfo{volume}{93}},
  \bibinfo{pages}{196805} (\bibinfo{year}{2004}{\natexlab{b}}).

\bibitem[{\citenamefont{Gambino and Colgan}(1998)}]{Gambino1998}
\bibinfo{author}{\bibfnamefont{J.}~\bibnamefont{Gambino}} \bibnamefont{and}
  \bibinfo{author}{\bibfnamefont{E.}~\bibnamefont{Colgan}},
  \bibinfo{journal}{Materials Chemistry and Physics}
  \textbf{\bibinfo{volume}{52}}, \bibinfo{pages}{99} (\bibinfo{year}{1998}),
  ISSN \bibinfo{issn}{0254-0584}.

\bibitem[{\citenamefont{Iwai et~al.}(2002)\citenamefont{Iwai, Ohguro, and
  Ohmi}}]{Iwai2002}
\bibinfo{author}{\bibfnamefont{H.}~\bibnamefont{Iwai}},
  \bibinfo{author}{\bibfnamefont{T.}~\bibnamefont{Ohguro}}, \bibnamefont{and}
  \bibinfo{author}{\bibfnamefont{S.-i.} \bibnamefont{Ohmi}},
  \bibinfo{journal}{Microelectronic Engineering} \textbf{\bibinfo{volume}{60}},
  \bibinfo{pages}{157} (\bibinfo{year}{2002}), ISSN \bibinfo{issn}{0167-9317}.

\bibitem[{\citenamefont{Morimoto et~al.}(1995)\citenamefont{Morimoto, Ohguro,
  Momose, Iinuma, Kunishima, Suguro, Katakabe, Nakajima, Tsuchiaki, Ono
  et~al.}}]{Morimoto1995}
\bibinfo{author}{\bibfnamefont{T.}~\bibnamefont{Morimoto}},
  \bibinfo{author}{\bibfnamefont{T.}~\bibnamefont{Ohguro}},
  \bibinfo{author}{\bibfnamefont{S.}~\bibnamefont{Momose}},
  \bibinfo{author}{\bibfnamefont{T.}~\bibnamefont{Iinuma}},
  \bibinfo{author}{\bibfnamefont{I.}~\bibnamefont{Kunishima}},
  \bibinfo{author}{\bibfnamefont{K.}~\bibnamefont{Suguro}},
  \bibinfo{author}{\bibfnamefont{I.}~\bibnamefont{Katakabe}},
  \bibinfo{author}{\bibfnamefont{H.}~\bibnamefont{Nakajima}},
  \bibinfo{author}{\bibfnamefont{M.}~\bibnamefont{Tsuchiaki}},
  \bibinfo{author}{\bibfnamefont{M.}~\bibnamefont{Ono}}, \bibnamefont{et~al.},
  \bibinfo{journal}{Electron Devices, IEEE Transactions on}
  \textbf{\bibinfo{volume}{42}}, \bibinfo{pages}{915} (\bibinfo{year}{1995}),
  ISSN \bibinfo{issn}{0018-9383}.

\bibitem[{\citenamefont{Murarka}(1995)}]{Murarka1995}
\bibinfo{author}{\bibfnamefont{S.~P.} \bibnamefont{Murarka}},
  \bibinfo{journal}{Intermetallics} \textbf{\bibinfo{volume}{3}},
  \bibinfo{pages}{173} (\bibinfo{year}{1995}), ISSN \bibinfo{issn}{0966-9795}.

\bibitem[{\citenamefont{Bucher et~al.}(1986-06-01)\citenamefont{Bucher, Schulz,
  Lux-Steiner, Munz, Gubler, and Greuter}}]{Bucher1986-06-01}
\bibinfo{author}{\bibfnamefont{E.}~\bibnamefont{Bucher}},
  \bibinfo{author}{\bibfnamefont{S.}~\bibnamefont{Schulz}},
  \bibinfo{author}{\bibfnamefont{M.~C.} \bibnamefont{Lux-Steiner}},
  \bibinfo{author}{\bibfnamefont{P.}~\bibnamefont{Munz}},
  \bibinfo{author}{\bibfnamefont{U.}~\bibnamefont{Gubler}}, \bibnamefont{and}
  \bibinfo{author}{\bibfnamefont{F.}~\bibnamefont{Greuter}},
  \emph{\bibinfo{title}{Work function and barrier heights of transition metal
  silicides}} (\bibinfo{year}{1986-06-01}).

\bibitem[{\citenamefont{Cai et~al.}(2012)\citenamefont{Cai, Che, Pelz,
  Hemesath, and Lauhon}}]{Cai2012a}
\bibinfo{author}{\bibfnamefont{W.}~\bibnamefont{Cai}},
  \bibinfo{author}{\bibfnamefont{Y.}~\bibnamefont{Che}},
  \bibinfo{author}{\bibfnamefont{J.~P.} \bibnamefont{Pelz}},
  \bibinfo{author}{\bibfnamefont{E.~R.} \bibnamefont{Hemesath}},
  \bibnamefont{and} \bibinfo{author}{\bibfnamefont{L.~J.}
  \bibnamefont{Lauhon}}, \bibinfo{journal}{Nano Lett.}
  \textbf{\bibinfo{volume}{12}}, \bibinfo{pages}{694} (\bibinfo{year}{2012}),
  ISSN \bibinfo{issn}{1530-6984}.

\bibitem[{\citenamefont{Lee et~al.}(2004)\citenamefont{Lee, Gipp, and
  Heller}}]{Lee2004}
\bibinfo{author}{\bibfnamefont{J.~U.} \bibnamefont{Lee}},
  \bibinfo{author}{\bibfnamefont{P.~P.} \bibnamefont{Gipp}}, \bibnamefont{and}
  \bibinfo{author}{\bibfnamefont{C.~M.} \bibnamefont{Heller}},
  \bibinfo{journal}{Appl. Phys. Lett.} \textbf{\bibinfo{volume}{85}},
  \bibinfo{pages}{145} (\bibinfo{year}{2004}).

\bibitem[{\citenamefont{Mueller et~al.}(2010)\citenamefont{Mueller, Kinoshita,
  Steiner, Perebeinos, Bol, Farmer, and Avouris}}]{Mueller2010}
\bibinfo{author}{\bibfnamefont{T.}~\bibnamefont{Mueller}},
  \bibinfo{author}{\bibfnamefont{M.}~\bibnamefont{Kinoshita}},
  \bibinfo{author}{\bibfnamefont{M.}~\bibnamefont{Steiner}},
  \bibinfo{author}{\bibfnamefont{V.}~\bibnamefont{Perebeinos}},
  \bibinfo{author}{\bibfnamefont{A.~A.} \bibnamefont{Bol}},
  \bibinfo{author}{\bibfnamefont{D.~B.} \bibnamefont{Farmer}},
  \bibnamefont{and} \bibinfo{author}{\bibfnamefont{P.}~\bibnamefont{Avouris}},
  \bibinfo{journal}{Nat Nano} \textbf{\bibinfo{volume}{5}}, \bibinfo{pages}{27}
  (\bibinfo{year}{2010}), ISSN \bibinfo{issn}{1748-3387}.

\bibitem[{\citenamefont{Heinzig et~al.}(2012)\citenamefont{Heinzig, Slesazeck,
  Kreupl, Mikolajick, and Weber}}]{Heinzig2011}
\bibinfo{author}{\bibfnamefont{A.}~\bibnamefont{Heinzig}},
  \bibinfo{author}{\bibfnamefont{S.}~\bibnamefont{Slesazeck}},
  \bibinfo{author}{\bibfnamefont{F.}~\bibnamefont{Kreupl}},
  \bibinfo{author}{\bibfnamefont{T.}~\bibnamefont{Mikolajick}},
  \bibnamefont{and} \bibinfo{author}{\bibfnamefont{W.~M.} \bibnamefont{Weber}},
  \bibinfo{journal}{Nano Lett.} \textbf{\bibinfo{volume}{12}},
  \bibinfo{pages}{119} (\bibinfo{year}{2012}), ISSN \bibinfo{issn}{1530-6984}.

\bibitem[{\citenamefont{Martin et~al.}(2011)\citenamefont{Martin, Heinzig,
  Grube, Geelhaar, Mikolajick, Riechert, and Weber}}]{Martin2011}
\bibinfo{author}{\bibfnamefont{D.}~\bibnamefont{Martin}},
  \bibinfo{author}{\bibfnamefont{A.}~\bibnamefont{Heinzig}},
  \bibinfo{author}{\bibfnamefont{M.}~\bibnamefont{Grube}},
  \bibinfo{author}{\bibfnamefont{L.}~\bibnamefont{Geelhaar}},
  \bibinfo{author}{\bibfnamefont{T.}~\bibnamefont{Mikolajick}},
  \bibinfo{author}{\bibfnamefont{H.}~\bibnamefont{Riechert}}, \bibnamefont{and}
  \bibinfo{author}{\bibfnamefont{W.~M.} \bibnamefont{Weber}},
  \bibinfo{journal}{Phys. Rev. Lett.} \textbf{\bibinfo{volume}{107}},
  \bibinfo{pages}{216807} (\bibinfo{year}{2011}).

\bibitem[{\citenamefont{Bachtold et~al.}(2001)\citenamefont{Bachtold, Hadley,
  Nakanishi, and Dekker}}]{Bachtold2001}
\bibinfo{author}{\bibfnamefont{A.}~\bibnamefont{Bachtold}},
  \bibinfo{author}{\bibfnamefont{P.}~\bibnamefont{Hadley}},
  \bibinfo{author}{\bibfnamefont{T.}~\bibnamefont{Nakanishi}},
  \bibnamefont{and} \bibinfo{author}{\bibfnamefont{C.}~\bibnamefont{Dekker}},
  \bibinfo{journal}{Science} \textbf{\bibinfo{volume}{294}},
  \bibinfo{pages}{1317} (\bibinfo{year}{2001}).

\bibitem[{\citenamefont{Derycke et~al.}(2001)\citenamefont{Derycke, Martel,
  Appenzeller, and Avouris}}]{Derycke2001}
\bibinfo{author}{\bibfnamefont{V.}~\bibnamefont{Derycke}},
  \bibinfo{author}{\bibfnamefont{R.}~\bibnamefont{Martel}},
  \bibinfo{author}{\bibfnamefont{J.}~\bibnamefont{Appenzeller}},
  \bibnamefont{and} \bibinfo{author}{\bibfnamefont{P.}~\bibnamefont{Avouris}},
  \bibinfo{journal}{Nano Letters} \textbf{\bibinfo{volume}{1}},
  \bibinfo{pages}{453} (\bibinfo{year}{2001}), ISSN \bibinfo{issn}{1530-6984}.

\bibitem[{\citenamefont{Huang et~al.}(2001)\citenamefont{Huang, Duan, Cui,
  Lauhon, Kim, and Lieber}}]{Huang2001}
\bibinfo{author}{\bibfnamefont{Y.}~\bibnamefont{Huang}},
  \bibinfo{author}{\bibfnamefont{X.}~\bibnamefont{Duan}},
  \bibinfo{author}{\bibfnamefont{Y.}~\bibnamefont{Cui}},
  \bibinfo{author}{\bibfnamefont{L.~J.} \bibnamefont{Lauhon}},
  \bibinfo{author}{\bibfnamefont{K.-H.} \bibnamefont{Kim}}, \bibnamefont{and}
  \bibinfo{author}{\bibfnamefont{C.~M.} \bibnamefont{Lieber}},
  \bibinfo{journal}{Science} \textbf{\bibinfo{volume}{294}},
  \bibinfo{pages}{1313} (\bibinfo{year}{2001}).

\bibitem[{\citenamefont{Chen et~al.}(2006)\citenamefont{Chen, Appenzeller, Lin,
  Sippel-Oakley, Rinzler, Tang, Wind, Solomon, and Avouris}}]{Chen2006}
\bibinfo{author}{\bibfnamefont{Z.}~\bibnamefont{Chen}},
  \bibinfo{author}{\bibfnamefont{J.}~\bibnamefont{Appenzeller}},
  \bibinfo{author}{\bibfnamefont{Y.-M.} \bibnamefont{Lin}},
  \bibinfo{author}{\bibfnamefont{J.}~\bibnamefont{Sippel-Oakley}},
  \bibinfo{author}{\bibfnamefont{A.~G.} \bibnamefont{Rinzler}},
  \bibinfo{author}{\bibfnamefont{J.}~\bibnamefont{Tang}},
  \bibinfo{author}{\bibfnamefont{S.~J.} \bibnamefont{Wind}},
  \bibinfo{author}{\bibfnamefont{P.~M.} \bibnamefont{Solomon}},
  \bibnamefont{and} \bibinfo{author}{\bibfnamefont{P.}~\bibnamefont{Avouris}},
  \bibinfo{journal}{Science} \textbf{\bibinfo{volume}{311}},
  \bibinfo{pages}{1735} (\bibinfo{year}{2006}).

\end{thebibliography}



\end{document}